\documentclass{article}
\usepackage{spconf,amsmath,graphicx}
\usepackage{math-common}
\usepackage{amsthm}
\usepackage{algorithm}
\usepackage{algpseudocode}%
\usepackage{booktabs}
\usepackage{color}
\usepackage{hyperref}

\title{High-Dimensional Sparse Bayesian Learning \\ without Covariance Matrices}
\name{Alexander Lin$^{\star}$ \qquad Andrew H. Song$^{\dagger}$ \qquad Berkin Bilgic$^{\dagger \ddagger \mathsection}$ \qquad Demba Ba$^{\star}$}
  
  \address{$^{\star}$School of Engineering and Applied Sciences, Harvard University, Boston, MA, USA \\
      $^{\dagger}$Massachusetts Institute of Technology, Cambridge, MA, USA \\ 
      $^{\ddagger}$Athinoula A. Martinos Center for Biomedical Imaging, Charlestown, MA, USA \\
      $^{\mathsection}$Department of Radiology, Harvard Medical School, Boston, MA, USA
     }

\begin{document}
%
\maketitle
\begin{abstract}
Sparse Bayesian learning (SBL) is a powerful framework for tackling the sparse coding problem.  However, the most popular inference algorithms for SBL become too expensive for high-dimensional settings, due to the need to store and compute a large covariance matrix.  We introduce a new inference scheme that avoids explicit construction of the covariance matrix by solving multiple linear systems in parallel to obtain the posterior moments for SBL.  Our approach couples a little-known diagonal estimation result from numerical linear algebra with the conjugate gradient algorithm.  On several simulations, our method scales better than existing approaches in computation time and memory, especially for structured dictionaries capable of fast matrix-vector multiplication.
\end{abstract}
\section{Introduction}
\label{sec:intro}

Sparse Bayesian learning (SBL) is {an effective tool} for \emph{sparse coding} -- the problem of identifying a small set of non-zero dictionary coefficients to explain the variance of large data.  It forms the basis for popular models, such as sparse Bayesian regression \cite{mackay1996bayesian}, relevance vector machines \cite{tipping2001sparse}, and Bayesian compressed sensing \cite{ji2008bayesian, ji2008multitask}.  It has found use in many applications, such as medical image reconstruction \cite{bilgic2011multi}, hyperspectral imaging \cite{ akhtar2015bayesian}, 
human pose estimation \cite{ babagholami2014bayesian}, and  structural health monitoring \cite{zhang2015sparse}.  Futhermore, SBL offers several advantages compared to other sparse coding approaches (e.g. $\ell_0$ regularization, $\ell_1$ regularization).  It provides uncertainty quantification \cite{tipping2001sparse}, removes the need to tune regularization penalties \cite{ji2008multitask}, exhibits favorable optimization properties \cite{wipf2007new}, enables active learning \cite{seeger2007bayesian}, and can be embedded as a submodule in complex generative frameworks \cite{fang2014pattern, wu2015space}.

However, one often-noted limitation of sparse Bayesian learning is the heavy computational cost of inference \cite{tipping2001sparse, bilgic2011multi}.  In terms of both time and space complexity, existing inference algorithms scale poorly to very high-dimensional problems.  Most algorithms have time complexity that is polynomial in $D$, the dimension of the signal to be recovered.  Unfortunately, in many practical settings, $D$ can be very large ($\geq 10^5$), leading to slow learning.  One way to accelerate SBL inference is to employ hardware optimized for parallel computing, such as graphics processing units (GPUs).  However, GPUs have limited memory, while most existing SBL algorithms require at least quadratic space, leading to memory issues for high $D$.   

We introduce a new approach to SBL inference that is more scalable than existing approaches for very large $D$.  We call our method \emph{covariance-free expectation-maximization (CoFEM)}, since it circumvents the main challenge of maintaining and inverting a $D \times D$ covariance matrix. This is possible by leveraging tools from numerical linear algebra, such as the diagonal estimation rule and conjugate gradient algorithm. CoFEM has $O(\tau_D)$-time complexity, where $\tau_D$ is the time required for matrix-vector multiplication. This can be as low as $\tau_D= O(D\log D)$ for structured matrices commonly used in signal processing (e.g. convolution, Fourier transform).  Furthermore, CoFEM has a space complexity of $O(D)$, which enables further acceleration via GPUs while ameliorating potential memory issues.  In practice, CoFEM can be up to thousands of times faster than existing baselines.\footnote{Our code can be found at \href{https://github.com/al5250/sparse-bayes-learn}{https://github.com/al5250/sparse-bayes-learn}}

\section{Background}
\label{sec:background}

\subsection{Generative Model}
The generative model for sparse Bayesian learning is
\begin{align}
\bd z &\sim \mathcal{N}(\bd 0, \text{diag}\{\bd \alpha\}^{-1}), \nonumber \\
\bd y &\sim \mathcal{N}(\bd \Phi \bd z, 1 / \beta\,\mathbf I), \label{model}
\end{align}
where $\bd z \in \R^D$ is a \emph{sparse latent vector}, $\bd y \in \R^N$ is an \emph{observation vector}, $\bd \Phi \in \R^{N \times D}$ is a known \emph{dictionary}, $\beta$ is the precision of the observation noise, and $\mathbf I$ is the $N \times N$ identity matrix.  Given $\bd y$, the goal of SBL inference is to recover $\bd z$.

The main feature of SBL is the diagonal Gaussian prior with precision parameters $\bd \alpha \in \R^D$ placed on $\bd z$.  SBL performs type II maximum likelihood estimation by integrating out $\bd z$ and optimizing $\bd \alpha$ \cite{wipf2011latent}, after which one can compute the posterior and recover sparse signals.  The learning objective is                       
\begin{align}
\max_{\bd \alpha}& \log p(\bd y \given \bd \alpha) = \log \int_{\bd z} p(\bd y \given \bd z) p(\bd z \given \bd \alpha) d \bd z  \label{mle2}.
\end{align}

During optimization, many elements of $\bd \alpha$ diverge to $\infty$. Consequently, the independent Gaussian priors over these elements of $\bd z$ converge to point masses on zero, forcing their respective posteriors to do the same. Thus, after $\bd \alpha$ converges, the posterior {$p(\bd z \given \bd y, \bd \alpha)$} is often highly sparse.

\subsection{Existing Inference Frameworks}
Inference schemes for SBL are designed to optimize Eq. \eqref{mle2}.
The most popular algorithm is \emph{expectation-maximization (EM)} \cite{tipping2001sparse}, which alternates between an E-Step and an M-Step to iteratively optimize Eq. \eqref{mle2}. Given an estimate $\bhat \alpha$, the E-Step computes the posterior $p(\bd z \given \bd y, \bhat \alpha)\sim \mathcal{N}(\bd \mu, \bd \Sigma)$ with
\begin{align}
\bd \mu = \beta \bd \Sigma \bd \Phi^\top \bd y, & & \bd \Sigma = (\beta \bd \Phi^\top \bd \Phi + \text{diag}\{\bhat \alpha\})^{-1}. \label{estep-params}
\end{align}  
Then, the M-Step uses these quantities to perform the update
\begin{align}
\bhat \alpha^\text{new} = \bd 1 \oslash (\bd \mu \odot \bd \mu + \bd \Sigma[\diagdown]), \label{mstep}
\end{align} 
where $\bd 1$ is the $D$-dimensional vector of ones, and $\odot$ and $\oslash$ denote element-wise multiplication and division, respectively.  The notation $\bd \Sigma[\diagdown]\in\mathbb{R}^D$ extracts the diagonal elements of $\bd \Sigma$.  
Despite its simplicity, EM suffers from heavy time and space complexity for large $D$. Specifically, the E-Step of Eq. \eqref{estep-params} requires $O(D^3)$ time and $O(D^2)$ space to compute $\bd \Sigma$.

Therefore, many existing SBL frameworks aim to accelerate EM; our proposed method also falls into this category. 
One line of work is \emph{iterative reweighted least-squares} (IRLS) \cite{wipf2010iterative}, which inverts an $N \times N$ covariance instead of a $D \times D$ one.  Although this can be faster when $N < D$, the time complexity remains a cubic function.   Another approach uses \emph{approximate message passing} (AMP).  Within each E-Step, AMP performs $T_\text{amp}$ iterative steps to approximate means and variances in Eq. \eqref{estep-params} to avoid matrix inversion \cite{fang2016two}.  However, AMP is known to diverge easily, especially for $\bd \Phi$ that do not satisfy zero-mean, sub-Gaussian criteria \cite{al2017gamp, luo2019sparse}.  AMP also requires computation of an $N \times D$ matrix $\bd \Phi \odot \bd \Phi$, which can be costly.  A third class of strategies uses \emph{variational inference} (VI), which approximates the true posterior $p(\bd z \given \bd y, \bd \alpha)$ with a simpler surrogate $q(\bd z)$ (e.g. independent Gaussian distributions) \cite{bishop2000variational, duan2017fast}.  The variational E-Step can thus be simplified 
and only requires $O(D)$-space. However, the drawback of VI is that it optimizes a lower bound on Eq. \eqref{mle2} instead of the true objective, leading to biased results. 

Finally, as an alternative to EM, there is an approach based on sequential optimization (Seq) of Eq. \eqref{mle2} \cite{tipping2003fast}.  Its complexities scale with $d$ -- the number of non-zero elements of $\bd z$ -- instead of $D$.  Thus, for truly sparse vectors with $d \ll D$, Seq can be faster than EM.  However, the algorithm's sequential nature limits the extent to which it can benefit from parallel computing, and it still requires storage of a covariance matrix.  

\section{COVARIANCE-FREE EM}
\label{sec:cofem}

We introduce covariance-free EM (CoFEM), which accelerates EM by obviating the need to invert or even compute the covariance matrix $\bd \Sigma$.  We leverage tools from the numerical linear algebra literature to accomplish this goal.  The main insight of CoFEM is that not all elements of $\bd \Sigma$ are required for the M-Step in Eq. \eqref{mstep}.  Indeed, we only need $\bd \mu$ and $\bd \Sigma[\diagdown]$ of the posterior to update $\bhat \alpha$.  We therefore propose a simplified E-Step that can estimate $\bd \mu$ and $\bd \Sigma[\diagdown]$ from \emph{solutions to linear systems}. First, we can re-express Eq. \eqref{estep-params} for $\bd \mu$ as 
\begin{align}
\bd \Sigma^{-1} \bd \mu = \beta \bd \Phi^\top \bd y,
\end{align}
where $\bd \Sigma^{-1} = \beta \bd \Phi^\top \bd \Phi + \text{diag}\{\bhat \alpha\}$.  Thus, $\bd \mu$ is the solution $\bd x$ to the linear system $\mathbf A \bd x = \bd b$ for $\mathbf A := \bd \Sigma^{-1}$ and $\bd b := \beta \bd \Phi^\top \bd y$.  Next, we estimate $\bd \Sigma[\diagdown]$ using the following result from \cite{bekas2007estimator}.

\subsection{Estimation of $\bd \Sigma[\diagdown]$}
\newtheorem*{prop}{Proposition}
\begin{prop}
[Diagonal Estimation Rule \cite{bekas2007estimator}] Let $\mathbf M$ be any square matrix of size $D \times D$.  Let $\bd p_1, \bd p_2, \ldots, \bd p_K \in \R^D$ be $K$ random probe vectors, where each $\bd p_k$ has independent and identically distributed components such that $\E[\bd p_k] = \bd 0$. Then the following $\bd s$ is an unbiased estimator of $\mathbf M[\diagdown]$,
\begin{align*}
\bd s = \left(\sum_{k=1}^K \bd p_k \odot \bold M \bd p_k\right) \oslash \left(\sum_{k=1}^K \bd p_k \odot \bd p_k \right).
\end{align*}
\end{prop}

We apply the diagonal estimation rule to $\bd \Sigma$ to estimate $\bd \Sigma[\diagdown]$.  We employ the \emph{Randemacher distribution}, to draw each component of $\bd p_k$ as either $+1$ or $-1$ with equal probability.  In this case, the diagonal estimator $\bd s$ simplifies to 
\begin{align}
\bd s = \frac{1}{K} \sum_{k=1}^K \bd p_k \odot \bold \Sigma \bd p_k, \label{diag-estimate}
\end{align}        
where $\E[\bd s] = \bold \Sigma[\diagdown]$. Eq. \eqref{diag-estimate} indicates that we need to apply $\bd \Sigma$ to each probe vector $\bd p_k$.  We can compute $\bd \Sigma \bd p_k$ by solving a linear system $\bold A \bd x = \bd b$ for $\bd x$, where $\bold A := \bd \Sigma^{-1}$ and $\bd b := \bd p_k$.  

In summary, $\bd \mu$ and $\bd \Sigma[\diagdown]$ can be obtained by solving $K + 1$ separate linear systems.  These systems can be solved \emph{in parallel} by considering the matrix equation $\bold A \bold X = \bold B$ with  
\begin{align}
\bold A &:= \beta \bd \Phi^\top \bd \Phi + \text{diag}\{\bhat \alpha\}, \nonumber \\ \bold B &:= 
\begin{bmatrix}
\bd p_1 \given \bd p_2 \given \ldots \given \bd p_K \given \beta \bd \Phi^\top \bd y \label{mat-eq}
\end{bmatrix}.
\end{align}  
If we enumerate the columns of the solution matrix $\bold X \in \R^{D \times (K + 1)}$ as $\bd x_1, \bd x_2, \ldots, \bd x_K, \bd \mu$, our desired quantities for the simplified E-Step are $\bd \mu$ and $\bd s := 1/K \sum_{k=1}^K \bd p_k \odot \bd x_k$.  We can then perform the M-Step of Eq. \eqref{mstep} as
\begin{align}
\bhat \alpha^\text{new} = \bd 1 \oslash (\bd \mu \odot \bd \mu + \bd s),
\end{align}
avoiding the need to compute or invert $\bd \Sigma$.  Algorithm \ref{cofem} summarizes the full CoFEM algorithm.
Preliminary theoretical analysis suggests that the variance of Eq. \eqref{diag-estimate} scales with $1 / K$ \cite{bekas2007estimator}.  In practice, we have found that small $K$ (e.g. $K = 20$) is sufficient even at high $D$ (Section \ref{sec:experiments}).  We defer a detailed theoretical discussion of the estimator to future work.

\subsection{Linear Solver for CoFEM}

Among potential options for the linear solver in Algorithm \ref{cofem}, we use conjugate gradient (CG) for several reasons \cite{hestenes1952methods, shewchuk1994introduction}.  CG does not require construction of the matrix $\bold A$ to solve $\bold A \bd x = \bd b$; we just need a way to apply $\bold A$ to an arbitrary vector $\bd v$.  Since our SBL framework defines $\bold A := \bd \Phi^\top \bd \Phi + \text{diag}(\bhat \alpha)$, the time complexity of CG (and CoFEM) scales according to $O(\tau_D)$, the time it takes to apply $\bd \Phi$ (and $\bd \Phi^\top$) to $\bd v$. For many structured matrices used in signal processing (e.g. discrete cosine transform, Fourier transform, wavelet transform, convolution), we have $O(\tau_D)=O(D \log D)$.  In addition, CG is space-efficient and only needs $O(D)$-space to solve the linear system; this is the minimum requirement for \emph{any} solver given that the output $\bd x\in\mathbb{R}^D$.  Furthermore, CG easily generalizes to multiple linear systems $\bold A \bold X = \bold B$ by simply replacing the matrix-vector multiplications with matrix-matrix multiplications. For accelerated computing, these operations can be parallelized on GPUs.  Finally, CG is an iterative approach that guarantees convergence to a solution within $\leq D$ steps. In practice, far fewer steps are needed to find an $\hat{\bold X}$ such that $\norm{\bold A \hat{\bold X} - \bold{B}}_F / \norm{\bold{B}}_F < \epsilon$ for small $\epsilon$, where $\norm{\cdot}_F$ denotes Frobenius norm \cite{shewchuk1994introduction}.  Thus, we can set an upper limit $U \ll D$ on the number of iterations and still obtain good performance. 


\subsection{Complexity Comparison}
Each of the $T_\text{em}$ iterations of CoFEM requires at most $U$ steps of CG -- in which we apply $\bd \Phi$ (and $\bd \Phi^\top$) in $O(\tau_D)$-time to $K$ vectors -- giving us an overall time complexity of $O(T_\text{em} \tau_D U K)$.  CoFEM's space complexity is dominated by CG, which requires $O(D)$-space for each of the $(K+1)$ systems. Table \ref{tab:complex} shows the complexities of CoFEM and other SBL inference schemes. While many other methods improve upon EM, they introduce dependencies on $N$ or $d$, which typically grow with $D$.  For example, if the size of the signal $\bd z$ is doubled, we may also expect the number of measurements $N$ to be doubled (to achieve same reconstruction error), as well as the number $d$ of non-zero values  in $\bd z$. Thus, increasing $D$ compounds the increase in complexities of these algorithms.  In contrast, CoFEM's dependencies on $U$ and $K$ can be held constant as $D$ increases, which we demonstrate in Section~\ref{sec:experiments}.

\section{EXPERIMENTS}
\label{sec:experiments}

We run a set of experiments on simulated data to compare CoFEM against other SBL inference methods, following the compressed sensing setup~\cite{ji2008bayesian}.  We focus on two different types of dictionaries -- \emph{dense} and \emph{structured}.

\begin{figure*}
    \centering
    \includegraphics[scale=0.5]{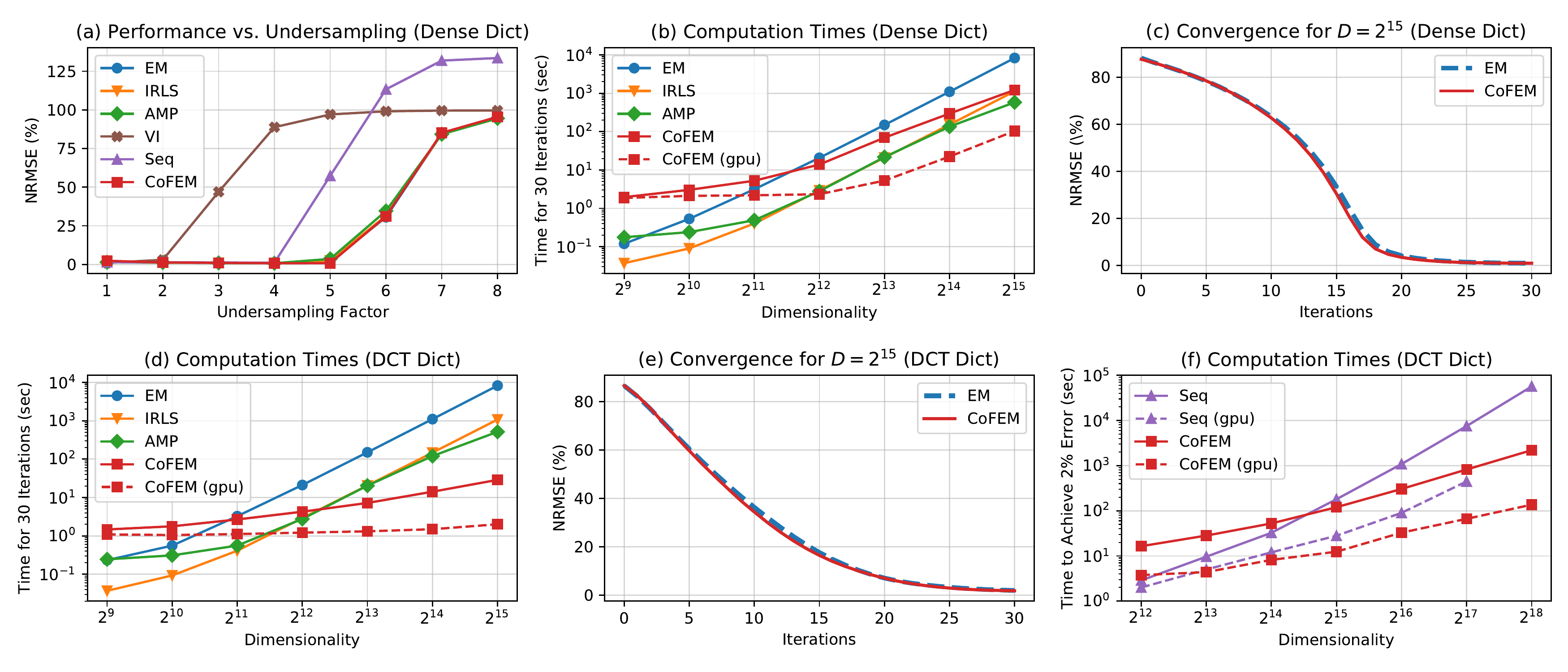}
    \caption{{Comparison of NRMSE and computation time between CoFEM and other SBL inference algorithms.}}
    \label{fig:main}
\end{figure*}

\subsection{Dense Dictionary}

\noindent \textbf{Recovery Accuracy} For various undersampling factors $f \in \{1, 2, \ldots, 8\}$, we generate a ground-truth sparse signal $\bd z^* \in \R^{D = 1024}$ and simulate undersampled data $\bd y = \bold \Phi \bd z^* + \bd \varepsilon$ of length $N = D / f$, 
where $\bd z^*$ has $d = 0.04 D$ randomly chosen spikes drawn from $\{-1, +1\}$ with equal probability (all other components are zero), $\bd \Phi \in \R^{N \times D}$ is a dense matrix drawn from $\mathcal{N}(0, 1)$, and $\bd \varepsilon \in \R^{N}$ is drawn from $\mathcal{N}(0, \sigma^2)$ with $\sigma = 0.005$ \cite{ji2008bayesian}.  Given $\bd y$ and $\bold \Phi$, each SBL inference algorithm outputs a reconstruction $\bd \mu$ after convergence.  We use normalized root mean square error (NRMSE) $\norm{\bd \mu - \bd z^*}_2 / \norm{\bd z^*}_2 \times 100\%$ to determine performance (lower is better).  We set $T_\text{em} = 50$ EM steps, $T_\text{amp} = 10$ inner AMP steps, $U = 400$ maximum CG iterations, and $K = 20$ probe vectors.  Figure \ref{fig:main}(a) displays NRMSE vs. $f$ for various algorithms (averaged across 25 repetitions).  We note that four methods -- EM, IRLS, AMP, and CoFEM -- have similar performance.  On the other hand, VI performs poorly due to optimization of a biased objective.  The sequential algorithm also deteriorates quickly for $f > 4$.

\begin{algorithm}[!t!]
\caption{\textsc{CovarianceFreeEM}($\bd y$, $\bd \Phi$, $\beta$, $T_\text{em}$, $K$)} \label{cofem}
\begin{algorithmic}[1]
\State{Initialize $\bhat \alpha \gets \bd 1$.}
\For {$t = 1, 2, \ldots, T_\text{em}$}
    \State{\emph{// Simplified E-Step }}
    \State{Define $\bold A \gets \beta \bd \Phi^\top \bd \Phi + \text{diag}\{\bhat \alpha\}$.}
    \State{Draw $\bd p_1, \bd p_2, \ldots, \bd p_K \sim $ Randemacher distribution.}
    \State{Define $\bold B \gets [\bd p_1 \given \bd p_2 \given \ldots \given \bd p_K \given \beta \bd \Phi^\top \bd y]$.}
    \State{$[\bd x_1 \given \bd x_2 \given \ldots \given \bd x_K \given \bd \mu] \gets $ \textsc{LinearSolver}($\bold A, \bold B$).}
    \State{Compute $\bd s \gets 1 / K\sum_{k=1}^K \bd p_k \odot \bd x_k$.}
    \State{\emph{// M-Step}}
    \State {Update $\bhat \alpha \gets \bd 1 \oslash (\bd \mu \odot \bd \mu + \bd s)$.}
\EndFor \\
\Return {$ \bhat \alpha, \bd \mu, \bd s$}
\end{algorithmic}
\end{algorithm}

\begin{table}
\vspace{-0.8em}
\centering
\caption{Comparing different SBL inference schemes.}\label{tab:complex}
\begin{tabular}{lcc}
\toprule
\textbf{Method} & \textbf{Time} & \textbf{Space}  \\ 
\hline
EM \cite{tipping2001sparse} & $O(T_\text{em} D^3)$ & $O(D^2)$ \\
IRLS \cite{wipf2010iterative} & $O(T_\text{em}(DN^2 + N^3))$ & $O(D^2)$\\
AMP \cite{fang2016two} & $O(T_\text{em} T_\text{amp} DN)$ & $O(DN)$ \\
VI \cite{duan2017fast} & $O(T_\text{em} \tau_D)$ & $O(D)$ \\
Seq \cite{tipping2003fast} & $O(Dd^2)$ & $O(D + d^2)$ \\
CoFEM (ours) & $O(T_\text{em} \tau_D U K)$ & $O(DK)$ \\
\bottomrule
\end{tabular}
\end{table}

\noindent \textbf{Computation Time} We compare computation time among the EM-based algorithms (EM, IRLS, AMP, CoFEM) for increasing $D$.  We exclude VI due to poor signal recovery performance and Seq due to different optimization procedure.  We fix $f = 4$ and vary $D = 2^p$ for $p \in \{9, 10, \ldots, 15\}$.  For CoFEM, we fix $U = 400$ and $K = 20$ for all $D$. 

Figure \ref{fig:main}(b) presents a log-log plot of the running times for 30 EM iterations, using the dense matrix $\bd \Phi$. We observe that CoFEM is faster than EM for large $D$. We also observe that IRLS and AMP take slightly less time for the values of $D$ we consider, yet the gap with CoFEM closes for large $D$; we attribute this to all three algorithms requiring $O(DN)$-time for dense $\bd \Phi$, since $\tau_D = O(DN)$ (Table \ref{tab:complex}).  We also include the performance of CoFEM on a GPU\footnote{We use a Nvidia T4 GPU with 16 GB RAM.} to illustrate further accelerations made possible by its low space complexity.  The other algorithms are not executable at high dimensions (i.e. $D = 2^{15}$) on our GPU due to their memory requirements.  Figure \ref{fig:main}(c) compares the convergence of CoFEM and EM over iterations for $D=2^{15}$ (the results for other $D$ are similar).  CoFEM converges at the same rate as EM even though CoFEM is much faster to execute. This also reflects how well the estimated diagonal $\bd s$ approximates the true $\bd \Sigma[\diagdown]$.

\subsection{Structured Dictionary}
A dense $\bd \Phi$ is the worst case in terms of time complexity for CoFEM.  Thus, we also experiment with a more structured $\bd \Phi$ for which the benefits of CoFEM are more prominent.  Specifically, we let $\bd \Phi \in \R^{N \times D}$ be an inverse discrete cosine transform (DCT) followed by an undersampling mask to select $N = D / 4$ out of $D$ coordinates as observations.  The true signal $\bd z^*$ is a vector of DCT coefficients with $d = 0.04 D$ components drawn from $\mathcal{N}(0, 1)$ (all other components are zero).  We simulate the data as $\bd y = \bd \Phi \bd z^* + \bd \varepsilon$.  The SBL algorithms are tasked with recovering $\bd z^*$ from $\bd y$ and $\bd \Phi$.  For DCT, $\bd \Phi$ and $\bd \Phi^\top$ can be applied to a vector in $\tau_D = O(D \log D)$-time, making CoFEM much faster compared to the dense case.   

Figure \ref{fig:main}(d) presents a plot of computation times for various algorithms with increasing $D$. VI is omitted due to poor signal recovery.  We observe that EM, IRLS, and AMP have similar performance as in the dense case of Fig. \ref{fig:main}(b), because they do not exploit the structured form of $\bd \Phi$. In contrast, CoFEM, which takes advantage of the structured dictionary through CG, can be faster by several orders of magnitude for large $D$.  With GPU acceleration, CoFEM is up to thousands of times faster than EM.  Figure \ref{fig:main}(e) presents the convergence of CoFEM and EM for $D = 2^{15}$.

\noindent\textbf{CoFEM vs. Seq.}
Due to low space complexity, both CoFEM and Seq can handle very high-dimensional computation.  For these two algorithms, we compare the computation time required for attaining low NRMSE of $2\%$, since their iterations are not directly comparable.
We repeat the DCT experiment at higher dimensions $D = 2^p$ for $p = \{12, 13, \ldots, 18\}$ with more coefficients to recover (i.e. $d = 0.1D$).  In all cases, we fix $U = 400$ and $K = 20$.   Figure \ref{fig:main}(f) shows computation time as a function of $D$.  Note that $D = 2^{18}$ is still a realistic scenario (e.g. standard medical images with $512 \times 512$ pixels are this size~\cite{bilgic2011multi}). 
We observe that CoFEM is much faster than Seq on both CPU and GPU for large $D$.  For $D = 2^{18}$, the GPU does not have enough memory to execute Seq, since it requires storing a quadratically-growing covariance matrix.  CoFEM does not suffer from this issue and fully leverages the GPU to be up to hundreds of times faster than Seq.  

\section{CONCLUSION}
In this paper, we accelerated the EM algorithm for sparse Bayesian learning (SBL) by developing a covariance-free EM method (CoFEM) that avoids matrix inversion. We leveraged tools from numerical linear algebra to efficiently scale the algorithm to high-dimensional settings. As potential extensions, we can apply CoFEM to multi-task SBL \cite{ji2008multitask}, block-sparse SBL \cite{fang2014pattern}, and non-negative SBL \cite{nalci2018rectified}.

\clearpage

\bibliographystyle{IEEEbib}
\bibliography{refs}

\end{document}